\documentclass[12pt,preprint,flushrt]{aastex}

\newcommand{\kms}          {\mbox{${\rm km~s^{-1}}$}}
\newcommand{\cc}           {\mbox{${\rm cm^{-3}}$}}
\newcommand{\e}            {\mbox{$^{-1}$}}

\newcommand{\simgt}        {\gtrsim}
\newcommand{\simlt}        {\lesssim}
\def\cm2{\mbox{${\rm cm^{-2}}$}}
\def\h2{\mbox{${\rm H}_2$}}
\def\nh2{\mbox{$n_{\rm H_2}$}}
\def\Nh2{\mbox{$N_{{\rm H}_2}$}}
\def\Mh2{\mbox{$M_{{\rm H}_2}$}}
\def\farcs{\hbox{$.\!\!^{''}$}}

\def\pp{\noindent\hangindent 20pt\hangafter=1}

\def\startfigcap{\vspace*{2.0\baselineskip}\bgroup\leftskip 0.45in\rightskip 0.45in}
\def\endfigcap{\par\egroup\vspace*{2.0\baselineskip}}
\def\plotfiddle#1#2#3#4#5#6#7{\centering \leavevmode
\vbox to#2{\rule{0pt}{#2}}
\includegraphics{#1}}

\begin{document}

\title{High resolution imaging of CO outflows in OMC-2 and OMC-3}
\author{Jonathan P. Williams}
\affil{Institute for Astronomy, 2680 Woodlawn Drive, Honolulu, HI 96822}
\email{jpw@ifa.hawaii.edu}
\author{R. L. Plambeck}
\affil{Astronomy Department, University of California, Berkeley, CA 94720}
\email{rplambeck@astro.berkeley.edu}
\author{Mark H. Heyer}
\affil{Department of Astronomy, LGRT 619, University of Massachusetts,
710 North Pleasant Street, Amherst, MA 01003}
\email{heyer@astro.umass.edu}

\shorttitle{CO outflows in OMC-2 and OMC-3}
\shortauthors{Williams et al.}

\begin{abstract}
A large scale, high resolution map of CO(1--0) emission
toward the OMC-2 and OMC-3 star forming regions is presented.
The map is a mosaic of 46 fields using the Berkeley-Illinois-Maryland
Array (BIMA) and covers $\sim 10'\times 15'$ at $\sim 10''$ resolution.
These data are combined with singledish FCRAO observations and analyzed
to identify and determine the properties of nine protostellar outflows.
The BIMA data alone almost completely resolve out the cloud emission
at central velocities and only recover $1/20$ of the flux in the high
velocity gas showing that outflows are generally broadly dispersed over
$\sim 1'$ angular scales. All nine identified outflows
emanate from known Class 0 or borderline Class 0/I sources,
are associated with knots of shocked \h2\ emission,
and have short dynamical times.
It is suggested that only the youngest, most spatially compact,
and energetic outflows have been found and that
more distributed high velocity gas undetected by BIMA is due
to older outflows continuing through the Class I phase of protostellar
evolution. The mechanical energy injection rate into the cloud
is estimated to be $\sim 1.5~L_\odot$ which is comparable to the
turbulent energy dissipation rate. Outflows appear capable, therefore,
of sustaining cloud turbulence but a high star formation rate is
required implying a short cloud lifetime $\leq 5$~Myr.
\end{abstract}


\clearpage
\section{Introduction}
The accretion of gas onto a protostar is accompanied by
collimated jets observed in the optical and infrared
(Mundt \& Fried 1983; Schwartz et al. 1988)
and swept up molecular gas outflows observed at millimeter wavelengths
(Snell, Loren, \& Plambeck 1980).
The resulting transfer of momentum and kinetic energy into a protostar's
immediate surroundings may limit further accretion and thereby
determine the final stellar mass (Shu, Adams, \& Lizano 1987).
The combined effect of multiple outflows averaged over space and
time may also regulate future star formation on larger scales by
maintaining cloud turbulence (McKee 1989).
This paper addresses the latter issue by analyzing the energy injection
rate into a cloud by a collection of very young outflows and the spatial
distribution of high velocity gas.

Molecular outflows are most powerful in the earliest, Class 0,
stage of protostellar evolution (Bontemps et al. 1996).
A chain of such young protostars was found by
Chini et al. (1997; hereafter C97) through mapping dust continuum
emission at $1300~\mu$m along a ridge of dense molecular gas north of
the Orion nebula. Subsequent maps at $350~\mu$m and $850~\mu$m were
made by Lis et al. (1998) and Johnstone \& Bally (1999) respectively.
The ridge is split into two main groups, OMC-2 in the southern half
and OMC-3 in the north. The ratio of sub-millimeter to far-infrared
luminosity is higher in OMC-3 than in OMC-2 suggesting a trend toward
younger protostars with increasing distance from the Orion nebula
(C97; Lis et al. 1998). However, all the sources found by C97 are deeply
embedded and can be classified as either Class 0 or borderline Class 0/I.
Other papers that are pertinent to this work are $2.12~\mu$m
imaging of the $\nu=1-0$ S(1) line of shock excited \h2\ 
by Yu, Bally, \& Devine (1997; hereafter Y97) and, more recently,
Stanke, McCaughrean, \& Zinnecker (2002; S02).
Y97 found 80 knots of \h2\ emission which they connected into
twelve jets from young protostars. The positions of these
knots are used as a guide to deciphering the molecular outflows
in the maps presented here. S02 mapped a much larger region in
Orion A and confirmed the positions of the Y97 knots in OMC2/3.
Yu et al. (2000; hereafter Y00) followed up their \h2\ imaging
study with molecular line observations of several jets.
C97 also presented molecular line observations. In both cases,
the relatively large beamsizes ($> 24''$) of these singledish
studies together with the high density of protostars and multiple,
overlapping outflows prevented a firm association of molecular
outflow gas with \h2\ jets. Aso et al. (2000; hereafter A00)
mapped CO(1--0) and denser gas tracers with the largest singledish
telescope operating at 3~mm, the Nobeyama 45~m, to achieve
maps of the outflows at $17''$. They found eight outflows
and their map is the most comparable to this work.

Because of the high protostellar density in the region,
it would be desirable to map the outflows at comparable
resolution to the dust continuum maps that identified
the young protostars.
Only interferometers can achieve such a resolution, $\sim 10''$,
in the low frequency lines that the outflows generally excite.
The ten 6~m antennae that make up BIMA\footnotemark
\footnotetext{Operated with support from the National Science
Foundation under grants AST-9981308 to UC Berkeley, AST-9981363 to
U. Illinois, and AST-9981289 to U. Maryland.}
provide excellent $uv-$coverage and a large primary beam which result
in an efficient instrument for creating high resolution, large scale
images of the millimeter sky.
This paper presents a multi-field BIMA mosaic of the CO(1--0) emission
in the OMC2/3 region and shows the network of young molecular outflows
from the clustered protostars in unprecedented detail.
The following sections present the observational technique
used to produce the map, general results from the data and
a description of each of the nine individual outflows that
were identified. The paper concludes with a discussion of the
implications of this work for outflow lifetimes, energetics,
and cloud turbulence.

\section{Observations}
A 46-field mosaic was observed with BIMA four times
from October 1998 to March 1999, twice each in the compact C and D
configurations. Baselines ranged from 6~m to 81~m.
The total on-source integration time was 14 hours, or slightly less
than 20 minutes per field.
Amplitude and phase were calibrated using 4 minute observations of
0530+135 interleaved with each 23 minute integration on source
(30 seconds per field).
Observations of the bright quasar 3C\,454.3 at the start of each
track showed the passband to be flat and no correction was applied.
The flux density scale was set by observing Mars during
the 1998 observations resulting in a flux for 0530+135 of 2.0~Jy.
The CO line was placed in the upper sideband centered in a 256 channel
window with 25~MHz bandwidth with a corresponding velocity resolution
of 0.25~\kms.
The total continuum bandwidth was 600~MHz but the integration time
per field was too short to obtain more than a noisy detection of the
brightest sources in the C97 map.
The data were reduced in the MIRIAD software package (Wright \& Sault 1993)
and the final CLEANed map had a spatial resolution of
$13\farcs 4\times 8\farcs 8$ at a position angle of $1.9^\circ$
and a (boxcar smoothed) velocity resolution of 0.5~\kms.
The final map covers $\sim 10'\times 15'$ at a spatial dynamic range,
$A_{\rm map}/A_{\rm beam}=5800$.

The 46 pointing centers were arranged in a hexagonal grid spaced by
$\Delta\alpha=50'', \Delta\delta=86''$. This resulted in a separation
between pointing centers of $100''$, about the same as the primary
beam of the 6~m antennae. Such a beam sampled mosaic is sensitive to
compact structures and is suited to mapping large areas with minimal
observing overhead. A comparison between a beam sampled and Nyquist
sampled mosaic of CO outflows in NGC 1333 showed negligible differences
(Plambeck \& Engargiola 2000). The noise level, 0.6~K ($1\sigma$) per
0.5~\kms\ velocity channel, was nearly uniform across the final map 
with a sharp falloff in sensitivity from 80\% to 30\% over $30''$
at the map borders.

Singledish data were taken at the Five College Radio Astronomy Observatory
14~m telescope (FCRAO\footnotemark
\footnotetext{FCRAO is supported in part by the
National Science Foundation under grant AST-0100793 and is operated
with permission of the Metropolitan District Commission, Commonwealth
of Massachusetts}) in March 1999 with the 16 element, focal plane
array SEQUOIA and a system of autocorrelation spectrometers at a
spectral resolution of 78~kHz. The observing was carried out using
standard position-switching procedures resulting in a Nyquist sampled
map with an rms noise of 0.3~K per 0.5~\kms\ channel.
Antenna temperatures were converted to fluxes using a gain of
44.5~Jy~K\e\ and added to the BIMA data to fill in the short
spacing information missing from the interferometer map through
joint maximum entropy deconvolution.

The addition of the singledish data partially compensates for
the spatial filtering properties of the interferometer and
makes possible flux recovery of extended objects.
Interferometers null out features that are uniform over angular sizes
$\theta\sim f\lambda/b_{\rm min}\simeq 90''f$
where $b_{\rm min}=6~{\rm m}=2.3~{\rm k}\lambda$ is the shortest
projected baseline separation and $f$ is a factor that theoretical
considerations suggest may be $\simeq 1/2$ (Wilner \& Welch 1994).
However, simulations by Helfer et al. (2002) show that the non-linear
deconvolution process can effectively extrapolate structural information
to larger values of $f$. Experience suggests that D-array observations
are sensitive to angular sizes $\simlt 60''$, i.e. $f=2/3$.
Since the beamsize of the FCRAO 14~m at 115~GHz is $47''$, the
combined BIMA+FCRAO dataset provides information on small scale
and large scale structures but poorly represents features
on angular scales in the range $\sim 60''-95''$.

\section{Results}
\subsection{Overview}
Average spectra over the mapped region for each of the two datasets are
shown in Figure 1. In contrast to the singledish map, large areas of
the interferometer map have very little emission so the average intensity
is much lower. Spectral profiles are also very different; the FCRAO
spectrum is centrally peaked with weak extended line-wing emission,
but the BIMA spectrum has a central dip and peaks in the line-wings.
This is due to the spatial filtering properties of interferometers
as described above. In this case, it can be concluded that the bulk
of the cloud emission at $v\simeq 9$ to 12.5~\kms\ is relatively smooth
over angular scales $\simgt 1'$ but that the line-wing emission
arises, at least in part, from more compact spatial features
(i.e. collimated outflows). Similar singledish/interferometer profiles
were observed toward NGC 1333 and HH 7-11 by Plambeck \& Engargiola (2000).

The BIMA data accounts for only 0.5\% of the flux of the total
cloud emission, as measured from the FCRAO data, demonstrating how
smooth the CO emission is. Even in the line-wings, however, the BIMA
data only accounts for 4.7\% of the emission so a large fraction of
the outflow emission must also be spatially extended over
$\sim 1'$ angular scales.
For this reason outflow properties are calculated from the combined
BIMA+FCRAO dataset that contains the flux of the singledish map.
The broad spatial extent of the high velocity
gas has important implications on its origin: either outflows can
rapidly distribute their momenta and energy over large regions
or most of the line-wing emission arises from a cascade of larger
scale processes such as Galactic shear (Fleck 1981).

Although the BIMA data accounts for only $\sim 1/20$ of the flux
of the line-wing emission, the spatial filtering properties of the
interferometer are advantageous in isolating the small scale features
of collimated outflows. Low and high velocity BIMA emission is plotted
in Figure 2. Several linear features can be seen and some clear outflows,
MMS5 and MMS10 being the best examples, with separated blue-red
lobes are apparent. Nine outflows were found in these data and they
are shown schematically in the map. All appear to originate from
one of the C97 protostars. The Y97 \h2\ knots are also plotted
in the Figure and were an essential guide in the assignment of the
line-wing emission to different outflows. Most \h2\ knots were found
to be associated with CO line-wing emission and those that were not
appear to be due to outflows in the plane of the sky. Similarly
most of the CO features in Figure 2 can be assigned to an outflow
from a C97 protostar but there are also many clumps of high velocity
gas whose origin remains unclear.

\subsection{Individual flows}
The identification of each of the nine outflows was made from analysis
of channel maps and position-velocity slices in conjunction with
the positions of the Y97 and S02 \h2\ knots. Physical properties
are listed in Table 1\footnotemark
\footnotetext{A distance to the cloud of 450~pc is assumed
(Brown, de Geus \& de Zeeuw 1994).}
ordered by the C97 driving source where MMS stands for
"millimeter source" and FIR for "far-infrared".
For each outflow the mass, momentum and energy were determined
by first outlining the extent of the blue and red lobes in position
and velocity space using interactive cursor-based IDL routines and
then employing equations 6 and 9 of Cabrit \& Bertout (1990).
The combined BIMA+FCRAO dataset was used for the calculations
and, following Y00, CO line-wing emission was taken to be
optically thin and at an excitation temperature of 30~K.
Eyeball estimates of sizes and velocity gradients were made from
the channel maps and position-velocity slices shown in Figures 3--7.

\smallskip
\noindent{\it MMS2/3:}
This east-west outflow runs close to the plane of the sky (Figure 3).
There is a very weak red lobe coincident with a string of
\h2\ knots and a compact blue lobe host to a cluster of
\h2\ knots. Because of its orientation,
this outflow is one of the longest identified in these maps and 
only a very slight velocity gradient could be measured across it.
MMS2 and MMS3 are too close to unambiguously identify one over the
other as the driving source of this flow. A large clump of high 
velocity gas lies to the west that may be an expanding shell or
loosely collimated outflow from any of MMS2/3, 4, 5 or 6 but it
is not associated with \h2\ emission and could not be conclusively
connected to any source.

\smallskip
\noindent{\it MMS5:}
This is a small east-west outflow with blue and red peaks
straddling MMS5 along the same direction as a line of 3 \h2\
knots (Figure 3). The position velocity map (Figure 7) along this line
shows the second highest velocity gradient measured out of the nine
flows and suggests a near pole-on orientation.
Two additional \h2\ knots point to a north-south jet from MMS6
(see also S02) but the CO emission is too confused in this area to
definitively identify the corresponding outflow.

\smallskip
\noindent{\it MMS7:}
Also known as IRAS 05329-0505 this source powers a bright reflection
nebula and drives a well collimated, one-sided (west) optical jet HH294
(Reipurth, Bally, \& Devine 1997). Y97 found a lone \h2\ knot $\sim 2'$
to the east that they suggested was the end of a long flow from FIR1c.
The data here suggest that this knot is in fact associated with MMS7
through a CO filament that appears to be an outflow in the plane of
the sky (Figure 4).
The line-wing map in Figure 2 shows no high velocity gas associated with
MMS7 but careful inspection of channel maps revealed a linear feature
connecting the \h2\ emission in HH294 with the eastern knot.
This putative outflow is just distinguishable above a broad plateau
of cloud emission in a small range of velocities near the systemic motion
of the cloud. Its direction is slightly different from the HH294
optical jet but agrees with the position angle of a small 3.6~cm radio
jet observed with the VLA (Reipurth et al. 2003).
Because of its orientation in the plane of the sky and immersion in
the general cloud emission, no velocity gradient could be measured
along the flow and only conservative estimates to its size and mass
could be made.

\smallskip
\noindent{\it MMS8:}
The identification of this flow is the most uncertain of the nine.
Channel maps and position-velocity cuts hint at an outflow around
MMS8 that connects several Y97 \h2\ knots (Figure 5) but there is
confusion from high velocity CO emission around MMS9 and the dense
chain of \h2\ knots from the well collimated jet that it powers.
As originally noted by Y00, the powerful CO outflow around MMS8
discussed in C97 was misidentified due to erroneous plotting
of source positions and is a combination of flows originating
from MMS9 and MMS10.

\smallskip
\noindent{\it MMS9:}
A prominent chain of \h2\ knots reveal a long east-west jet
originating from MMS9 (Y97, S02). The BIMA data show associated,
collimated, high velocity $v>13$~\kms\ CO (Figure 5).
On the eastern side of the source, the association of both \h2\ knots
and gas is less clear but a long, high velocity collimated flow is
discernable and best seen in Figure 2. The flow occurs over a narrow
range of velocities and suggests an orientation in the plane of the sky,
consistent with its projected length, almost 1~pc, by far the longest
of the nine flows in Table 1.

Two compact cores of low velocity gas lie to the west of MMS9 and were
connected to the high velocity core north of MMS10 by both Y00 and A00.
However, the greater resolution of these data clearly demonstrate that
the \h2\ jet is associated with the linear high velocity core $\sim 30''$
further south. A more diffuse red lobe is also found toward the blue
cores which suggests that this feature may be a pole on flow from
an unidentified source or an expanding shell from an older outflow.

\smallskip
\noindent{\it MMS10:}
This is a strong, clearly defined outflow running northeast-southwest
(Figure 5). Y00 suggest that the red lobe here is due to
the outflow from MMS9. While there is likely to be some contribution
from that flow, the higher resolution of these data show a clear
relationship in space and velocity (Figure 7) between MMS10 and
the strong blue and red lobes on either side of it. The \h2\ knots
appear to be associated with the edges of the CO lobes rather than
directly lining up with a single jet from the source itself.
This is one of the more massive flows identified in the cloud.

\smallskip
\noindent{\it FIR1b/c:}
This outflow occurs in a highly confused area of the maps where the
source density is high and there are multiple, overlapping flows.
Because of the large numbers of high velocity CO clumps and scattered
\h2\ knots here the identification of this outflow is uncertain.
The evidence for an outflow is based on the blue and neighboring red
lobes on either side of FIR1c and FIR1b and \h2\ knots in close, but
not exact, proximity (Figures 5, 6). A position velocity cut along
a line through the two sources is unable to distinguish one over the
other as the driving source (Figure 7).

\smallskip
\noindent{\it FIR2:}
This is a very weak but high velocity flow running north-south.
The two lobes almost overlap spatially (Figure 6) but extend over
a wide range in velocity (Figure 7) resulting in an extremely
high velocity gradient and suggesting a pole-on orientation.
\h2\ knots are found in the vicinity of FIR2 but their association
with this flow is uncertain and they may well be an extension of the
powerful MMS10 outflow.

\smallskip
\noindent{\it FIR3:}
A compact chain of \h2\ knots and intense, collimated, high velocity
CO emission reveal strong outflow activity. This source is a binary
with $4''$ separation (Pendleton et al. 1986) and appears to drive
two criss-crossed flows.
Channel maps show two blue lobes and a single, long, red lobe (Figure 6).
The position-velocity map in Figure 7 shows the two blue lobes more clearly.
Their red lobe counterparts can be seen in the two lowest contour levels
of the merged high velocity emission in the same Figure.
The sum of these two flows inject the most mass, momentum, and energy
into the cloud of the nine flows discussed in this work.

In addition, a potential tenth flow is seen in the BIMA
map (dashed line in Figure 2). This connects a single \h2\ knot
in the southwest with a chain of high velocity clumps that appear
to connect back to FIR3. It was, however, not possible to distinguish
this flow clearly in the combined BIMA+FCRAO dataset due to its
length, lack of strong velocity gradient, and the confusion of
background cloud emission.

\bigskip
\section{Discussion}
The OMC-2 and OMC-3 regions are highly active star forming centers
with multiple, overlapping outflows. High resolution, large scale
CO mapping allows more secure identification of the flows and their
driving sources and makes possible the determination of fundamental
physical properties such as mass, momentum, and energy. In practice,
low and high velocity clumps are found to be widely distributed
across the map and, particularly for the flows in the plane of the sky,
it was essential to use the Y97 and S02 \h2\ maps to assign these
features to outflows driven by a given source.

The nine identified CO outflows line up with $\sim 50$ of the 60 Y97
\h2\ knots in the mapped region. There were no clearly identifiable
bipolar CO outflows that were not associated with detectable molecular
hydrogen shocks.
This is partly, but not completely, due to the way in which the high
velocity gas is assigned to outflows. The additional information
provided by the CO data, however, shows that the extremes
of some of the jets found by Y97 appear to be misidentified.
For example, in Y97 the \h2\ knots 74/77, 76 and 79 define the ends of
jets I, E and J respectively but the CO data show that these knots are more
likely to be associated with flows from closer sub-millimeter sources,
MMS10, MMS7 and either MMS9 or 10 respectively.
These jets are likely, therefore, to be significantly smaller than
previously thought.
On the other hand, the MMS2/3 and MMS9 flows that lie in the plane
of the sky appear to have a measurable effect on the CO emission over
a sizeable fraction of a parsec in length as has been well established
for optical jets (Reipurth \& Bally 2001).

The large scale effect of outflows may also be apparent in the
``missing flux'' in the BIMA map due to structures that are larger
than $\sim 1'$. Over 95\% of the line-wing emission is
resolved out by the interferometer demonstrating that the high
velocity gas is broadly distributed.
Similarly, a map of the CO line-width at $10''$ resolution made from
the combined dataset is generally featureless across the cloud away
from a few peaks toward individual sources such as FIR3 and some
\h2\ knots along the MMS9 flow.

The fact that the line-wing emission is widely distributed
complicates the definition of the boundaries of the outflows and
the measurement of their physical properties. Additional uncertainties
are the optical depth of the CO line-wing emission and the inclination
of the flow to our line of sight. Since there are no $^{13}$CO
observations at the same resolution, optically thin emission
was assumed, providing a lower limit to the mass and kinematic
estimates. No inclination correction was applied to the individual
flow momenta and energies listed in Table 1 since it is difficult
to determine accurately.
Assuming, however, a mean inclination of $45^\circ$
for the outflow sample, total values of the mass, momenta
and energy are $1.4~M_\odot$, $5.4~M_\odot$~\kms\ and
$2.9\times 10^{44}$~erg respectively. The variation in masses of
individual outflows is about a factor of five but the velocities
do not greatly differ and the range of momenta and energies is only
a factor of six. There is no clear correlation between outflow
properties and the C97 driving source mass or luminosity.

Velocity gradients, $dV/dR\simeq (V_{\rm red}-V_{\rm blue}))/L$,
measured by eyeball fits to the position-velocity diagrams in
Figure 7, are given in Table 1. Velocity gradients for the two long
outflows from MMS2/3 and MMS9 that lie nearly in the plane of the
sky could not be accurately measured.
The inverse of the gradient in \kms\ pc\e\ is approximately
the dynamical time in Myr. The median dynamical time in this sample
is $2\times 10^4$~yr which is similar to the statistical lifetime
of Class 0 protostars in the $\rho$ Oph cluster
(Andr\'e \& Montmerle 1994). However, the identification of
the outflows here suffers from a bias toward either high velocity
gradients or sharp, collimated features that stand out against the
general cloud emission. The former corresponds to short dynamical
timescales and the latter is a probable indication of youth
(Masson \& Chernin 1993). It is not necessarily surprising, then,
that all nine flows described in this paper appear to be driven by a
young Class 0, or borderline Class 0/I, protostar in the C97 catalog.

The dynamical time is only a lower limit to the age of an
outflow (Parker, Padman, \& Scott 1991) but is probably a good
estimator for the young outflows observed here (Masson \& Chernin 1993).
If only the very youngest outflows, $t_{\rm flow}\sim 2\times 10^4$~yr,
have been identified, but (less well collimated) outflow activity
continues throughout the Class I phase,
$t_{\rm I}\sim {\rm few}\times 10^5$~yr (Wilking et al. 1989),
then the broadly distributed line-wing emission that was resolved out
in the BIMA map can be accounted for by outflows alone. In principle,
they should be apparent in a combined singledish+interferometer map
such as the one presented in this paper, but higher signal-to-noise data
are required to pick them out from the general cloud background.
This paper has focused on the high velocity gas that can be
identified with shocked \h2\ emission but there are several
other high velocity clumps in Figure 2 that were not associated
with outflows. These unassigned clumps may be parts of older
outflow shells which have broadened and slowed sufficiently such that
the shocks with the ambient medium are too weak to excite the
\h2\ $2.12\mu$m line. Earlier single dish maps of CO in Orion
indeed found several large scale, slowly expanding shells that were
interpreted as resulting from young stellar outflows (Heyer et al. 1992).

It has long been suggested that outflows can maintain cloud turbulence
(e.g., Norman \& Silk 1980). The agreement between the ratio of Class 0 to
Class I lifetimes, and the flux ratio between collimated and broad
outflows, suggests an approximately uniform energy input into the cloud
that can be estimated from the observed CO outflows here,
$$L_{\rm flow}={E_{\rm flow}\over t_{\rm flow}}
  ={2.9\times 10^{44}f_{\rm CO}~{\rm erg}\over 2\times 10^4~{\rm yr}}
  \simeq 5\times 10^{32}f_{\rm CO}~{\rm erg~s}\e,$$
where $f_{\rm CO}$ is a factor that corrects for the assumption
of optically thin CO line-wing emission in the calculation of the
outflow masses and energies.
Cabrit \& Bertout (1990) show that $f_{\rm CO}$ lies in the range $5-10$.
The cloud loses energy through turbulent decay.
Simulations of magnetohydrodynamic cloud turbulence show
rapid dissipation with a timescale,
$$t_{\rm diss}=\left({3.9\kappa\over M_{\rm rms}}\right)t_{\rm ff},$$
where $\kappa=\lambda_{\rm d}/\lambda_{\rm J}$
is the driving wavelength of the turbulence in units
of the Jean's length, $M_{\rm rms}$ is the rms Mach number
of the turbulence, and $t_{\rm ff}=(3\pi/32G\rho)^{1/2}$
is the free-fall timescale (MacLow 1999).
The mass of the OMC-2/3 complex is $1100~M_\odot$ (Lis et al. 1998).
The average temperature and 1-d velocity dispersion can be measured
from the FCRAO spectrum in Figure~1 and are equal to 20~K and 1.5~\kms\
respectively. The equivalent radius, measured from the projected area,
is 0.9~pc, and implies an average number density, $\nh2=5\times 10^3$~\cc.
These numbers imply a total kinetic (turbulent) energy
$E_{\rm turb}=7.4\times 10^{46}$~erg, Mach number $M_{\rm rms}=5$,
and free-fall time $t_{\rm ff}=5.2\times 10^5$~yr.
Parameterizing by $\kappa$, the dissipation time is
$t_{\rm diss}=4.1\times 10^6\kappa$~yr, and  the
turbulent energy dissipation rate is,
$$L_{\rm turb}={E_{\rm turb}\over t_{\rm diss}}
  =6\times 10^{33}\kappa^{-1}~{\rm erg~s}\e.$$

The average cloud density and temperature imply a Jean's
length, $\lambda_{\rm J}=0.2$~pc. If outflows sustain the turbulence,
the driving wavelength can be no greater than their size
$\lambda_{\rm d}\leq \langle L\rangle$.
The average outflow length in Table~1, corrected for an average
inclination angle of $45^\circ$, is $\langle L\rangle=0.5$~pc.
Thus $\kappa=\lambda_{\rm d}/\lambda_{\rm J}\leq 2.5$.
Shorter wavelength ``harmonics'' may also be present, however, and
MacLow (1999) argues that the driving wavelength must be less than
the Jean's length, i.e. $\kappa<1$, for turbulence to support the cloud.
Taking $\kappa\simeq 1$ and the maximum CO optical depth correction
$f_{\rm CO}=10$, the data are consistent with an approximate balance
between outflow energy injection rate and turbulent energy dissipation
rate,
$$L_{\rm flow}\simeq L_{\rm turb}
  \simeq 6\times 10^{33}~{\rm erg~s}\e=1.5~L_\odot.$$

To maintain the cloud turbulence over longer timescales than an
individual protostar's outflow lifetime requires a star formation rate
$\sim 9~{\rm stars}/2\times 10^4~{\rm yr}\sim 2.3\times 10^{-4}~M_\odot$~yr\e\
for an average protostellar mass of $0.5~M_\odot$ based on the Scalo (1986)
IMF. At this rate the OMC-2/3 complex has enough mass to continue forming
low mass stars for $5\times 10^6\epsilon$~yr where
$\epsilon=M_{\ast,{\rm tot}}/M_{\rm cloud}<1$ is the overall star
formation efficiency. Even if the cloud ultimately converts most of its
mass to stars, therefore, the high rate of turbulent energy dissipation
restricts its lifetime to a few Myr.
This is shorter than predicted for more massive GMCs
(Blitz \& Shu 1980; Williams \& McKee 1997)
but consistent with the rapid cloud formation and evolution scenario
proposed by Hartmann, Ballesteros-Paredes \& Bergin (2001) and the
observed small age spread of protostars in nearby star forming regions
(Hartmann 2001).

\acknowledgments
JPW is supported by NSF grant AST-0134739 and thanks
Bo Reipurth and Joan Najita for helpful discussions.

\section{References}
\parskip=0pt
\bigskip

\pp Andr\'{e}, P., \& Montmerle, T. 1994, ApJ, 420, 837

\pp Aso, Y., Tatematsu, K., Sekimoto, Y., Nakano, T., Umemoto, T.
    Koyama, K., \& Yamamoto, S. 2000, ApJS, 131, 465 (A00)


\pp Blitz, L., \& Shu, F. H. 1980, ApJ, 238, 148

\pp Bontemps, S., Andre, P., Terebey, S., \& Cabrit, S. 1996, A\& A, 311, 858

\pp Brown, A. G. A., de Geus, E. J., \& de Zeeuw, P. T. 1994, A\& A, 289, 101

\pp Cabrit, S., \& Bertout, C. 1990, ApJ, 348, 530

\pp Chini, R., Reipurth, B., Ward-Thompson, D., Bally, J., Nyman, L.-A.,
    Sievers, A., \& Billawala, Y. 1997, ApJ, 474, L135 (C97)

\pp Fleck, R. C. 1981, ApJ, 246, L151

\pp Hartmann, L. 2001, AJ, 121, 1030

\pp Hartmann, L., Ballesteros-Paredes, J. \& Bergin, E. A. 2001, ApJ, 562, 852

\pp Helfer, T. T., Vogel, S. N., Lugten, J. B., \& Teuben, P. J.
    2002, PASP, 114, 350

\pp Heyer, M., Morgan, J., Schloerb, F. P., Snell, R. L., Goldsmith, P. F.
    1992, ApJ, 395, 99

\pp Johnstone D., \& Bally, J. 1999, ApJ, 510, L49



\pp Lis, D. C., Serabyn, E., Keene, J., Dowell, C. D., Benford, D. J.
    Phillips, T. G., Hunter, T. R., \& Wang, N. 1998, ApJ, 509, 299

\pp MacLow, M.-M. 1999, ApJ, 524, 169

\pp Masson, C. R., \& Chernin, L. M. 1993, ApJ, 414, 230

\pp McKee, C. F. 1989, ApJ, 345, 782

\pp Mundt, R., \& Fried, J. W. 1983, ApJ, 274, L83

\pp Norman, C., \& Silk, J. 1980, ApJ, 238, 158

\pp Parker, N. D., Padman, R., \& Scott, P. F. 1991, MNRAS, 252, 442

\pp Pendleton, Y., Werner, M. W., Capps, R., \& Lester, D. 1986, ApJ, 311, 360

\pp Plambeck, R. L., \& Engargiola, G. 2000,
    in {\it Imaging at Radio through Submillimeter Wavelengths},
    eds. J. G. Mangum \& S. J. E. Radford, ASP Conf. Series, 217, 354

\pp Reipurth, B., \& Bally, J. 2001, Ann. Rev. Astron. Astr., 39, 403

\pp Reipurth, B., Bally, J., \& Devine, D. 1997, AJ, 114, 2708

\pp Reipurth, B., Rodriguez, L. F., Anglada, G., \& Bally, J. 2003, in prep.

\pp Scalo, J. 1986, Fund. Cosmic Phys., 11, 1

\pp Schwartz, R. D.. Jennings, D. G., Williams, P. M., \& Cohen, M. 1988,
    ApJ, 334, L99

\pp Shu, F.H., Adams, F.C., \& Lizano, S. 1987, A.R.A\&A., 25, 23

\pp Snell, R. L., Loren, R. B., \& Plambeck, R. L. 1980, ApJ, 239, L17

\pp Stanke, T., McCaughrean, M. J., \& Zinnecker, H. 2002, A\&A, 392, 239 (S02)



\pp Wilking, B. A., Lada, C. J., \& Young, E. T. 1989, ApJ, 340, 823

\pp Williams, J. P., \& McKee, C. F. 1997, ApJ, 476, 166

\pp Wilner, D. J., \& Welch, W. J. 1994, ApJ, 427, 898

\pp Wright, M. C. H., \& Sault, R. J. 1993, ApJ, 402, 546

\pp Yu, K., Bally, J., \& Devine, D. 1997, ApJ, 485, L45 (Y97)

\pp Yu, K., Billawala, Y., Smith, M. D., Bally, J., \& Butner, H. M.
    2000, AJ, 120, 1974 (Y00)

\clearpage
\begin{figure}[ht]
\plotfiddle{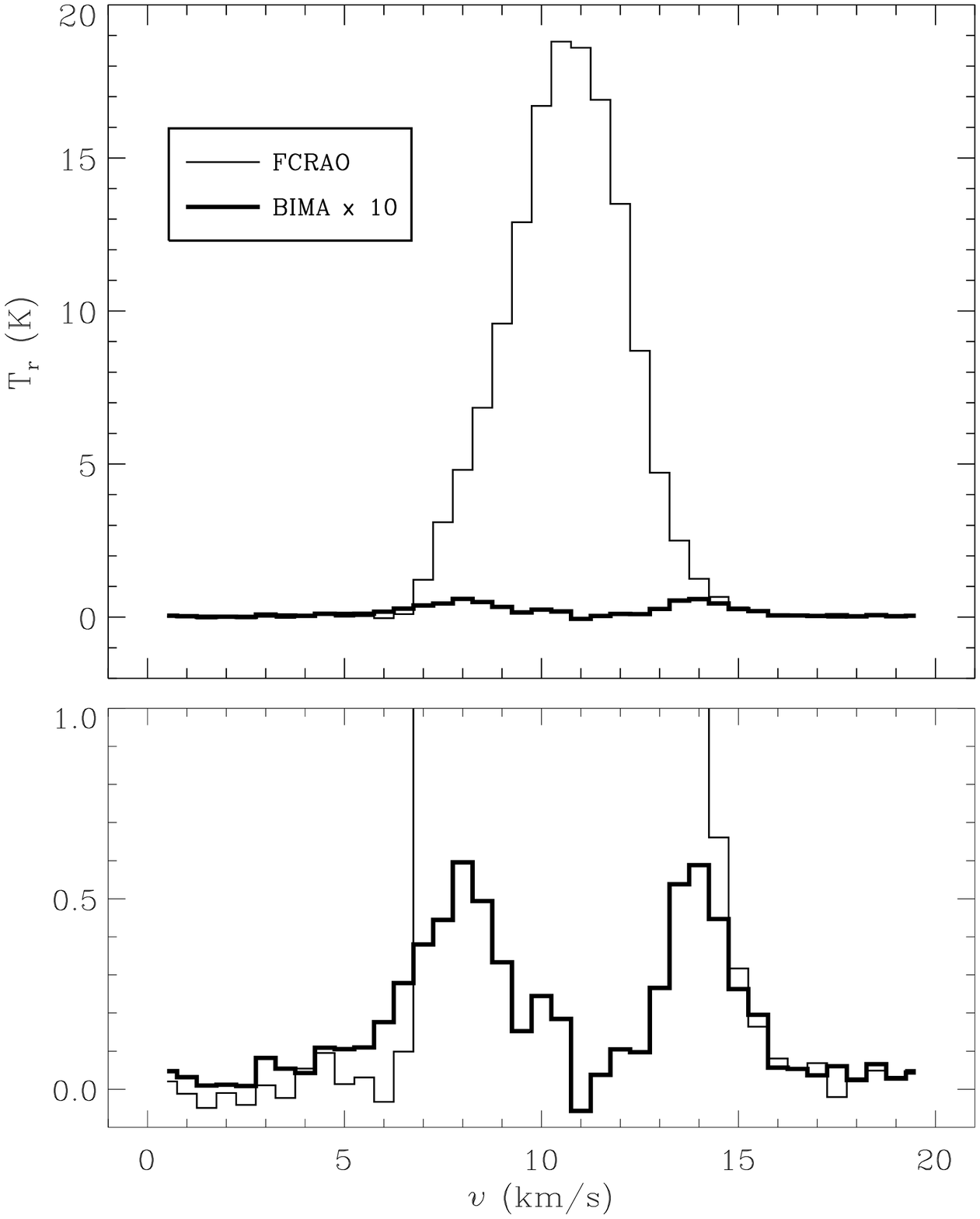}{50pt}{0}{60}{60}{-170}{-400}
\end{figure}
\vskip 4.8in
\startfigcap
\noindent{\bf Figure 1:}
Average spectra over the entire mapped region for the FCRAO and BIMA
data. The BIMA spectrum is scaled by a factor of 10 and plotted in the
heavier linetype. The lower panel zooms in on the lower intensity emission
and demonstrates that the interferometer data preferentially selects
the high and low velocity gas from the small scale features in outflows.
\endfigcap

\clearpage
\begin{figure}[ht]
\plotfiddle{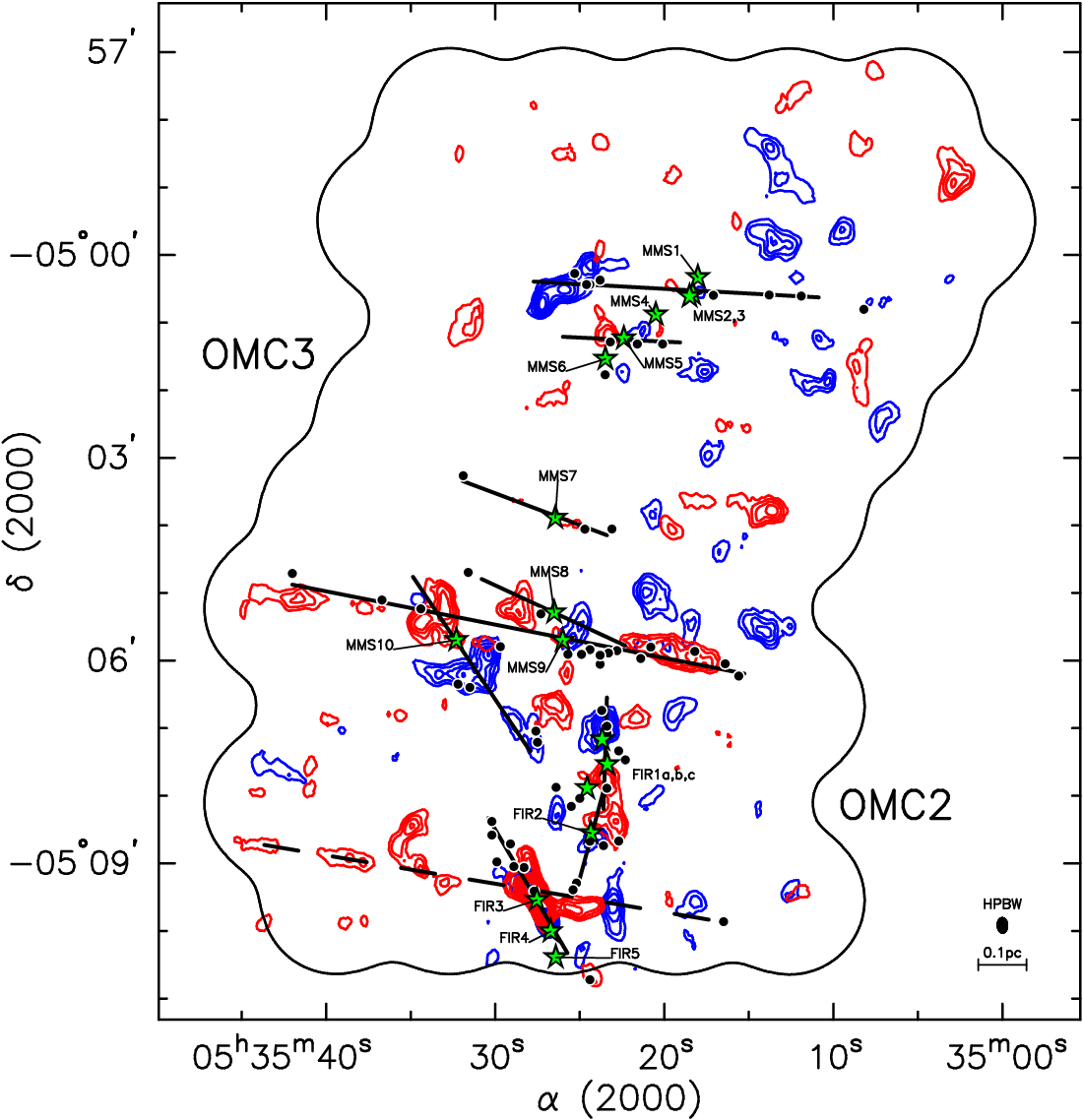}{50pt}{0}{80}{80}{-250}{-470}
\end{figure}
\vskip 4.4in
\startfigcap
\noindent{\bf Figure 2:}
Line-wing emission from the BIMA mosaicked map (interferometer
data only). Blue and red contours show the emission integrated from
$v=5.5$ to 9.0~\kms\ and 12.5 to 16.0~\kms\ respectively. Contour
starting levels and increments are the same (8, 4~K~\kms) for both
velocity ranges. The labeled star symbols show the location of the
C97 protostars and dots show the position of the Y97
$2.12~\mu$m \h2\ knots. Heavy dark lines show the outflows that
are discussed in this paper. The solid contour outlining the map
is the 30\% sensitivity level of the mosaic.
The $8.8''\times 13.4''$ beam is plotted in the lower right corner.
\endfigcap

\clearpage
\begin{figure}[ht]
\plotfiddle{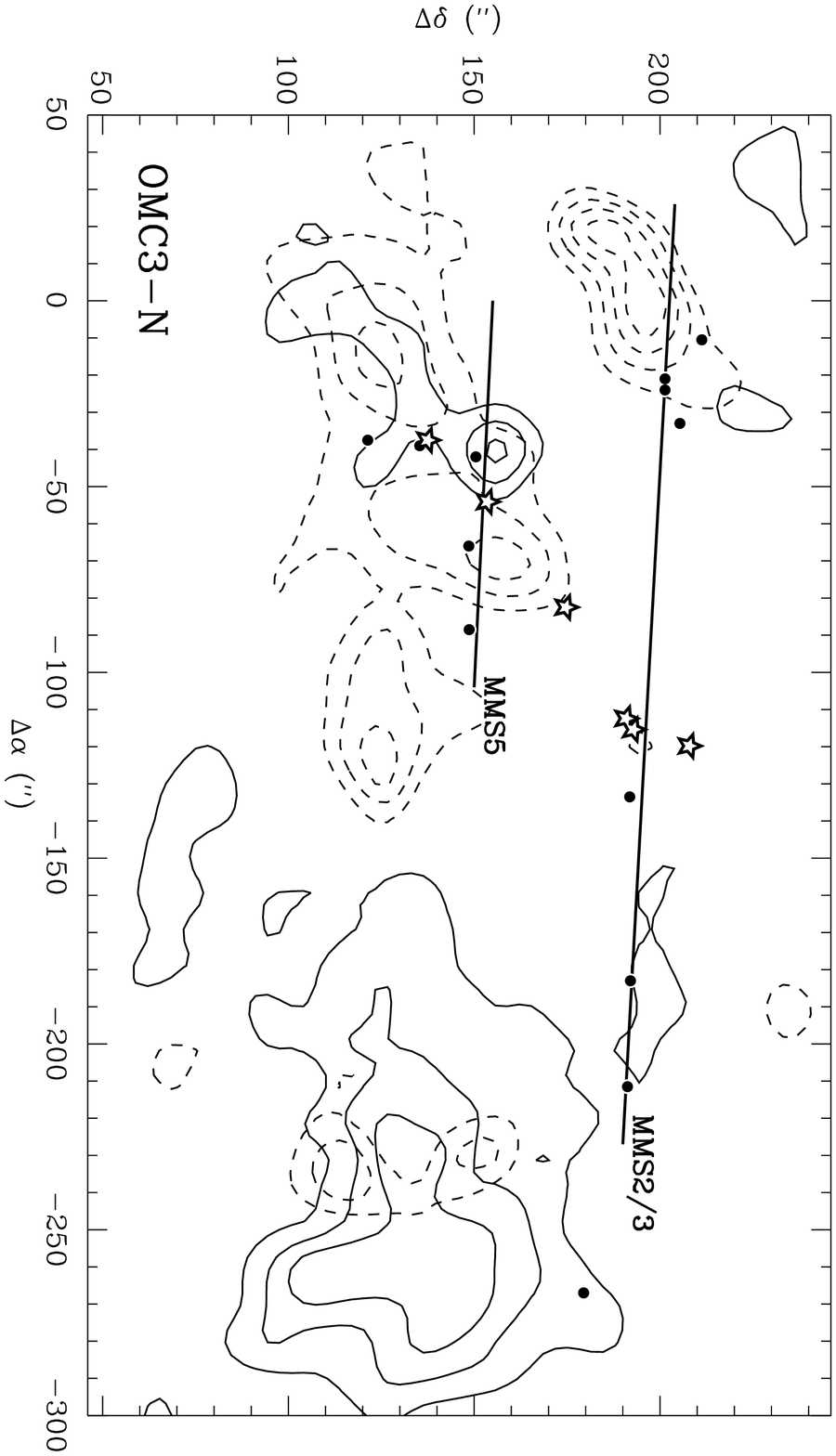}{50pt}{90}{60}{60}{240}{-300}
\end{figure}
\vskip 3.2in
\startfigcap
\noindent{\bf Figure 3:}
Line-wing emission from the combined BIMA+FCRAO mosaicked map
toward the northern OMC-3 region containing sources MMS1-6.
Coordinates are offset from $\alpha=5^{\rm h} 35^{\rm m} 26^{\rm s}$
and $\delta=-5^\circ 03^\prime 47.7^{\prime\prime}$ (J2000).
The dashed contours (colored blue in electronic edition)
show the low velocity gas integrated from
$v=5.0$ to 7.5~\kms\ and solid contours (colored red in electronic
edition) the high velocity gas
integrated from $v=13.5$ to 16.5~\kms. Contour starting levels
and increments are the same (8, 4~K~\kms) for both velocity ranges.
As in Figure 2, the star symbols show the location of the C97 protostars
and dots show the position of the Y97 $2.12~\mu$m \h2\ knots.
The labeled heavy solid lines show the outflows that are discussed in this
paper and along which the position-velocity maps are defined in Figure~7.
\endfigcap

\clearpage
\begin{figure}[ht]
\plotfiddle{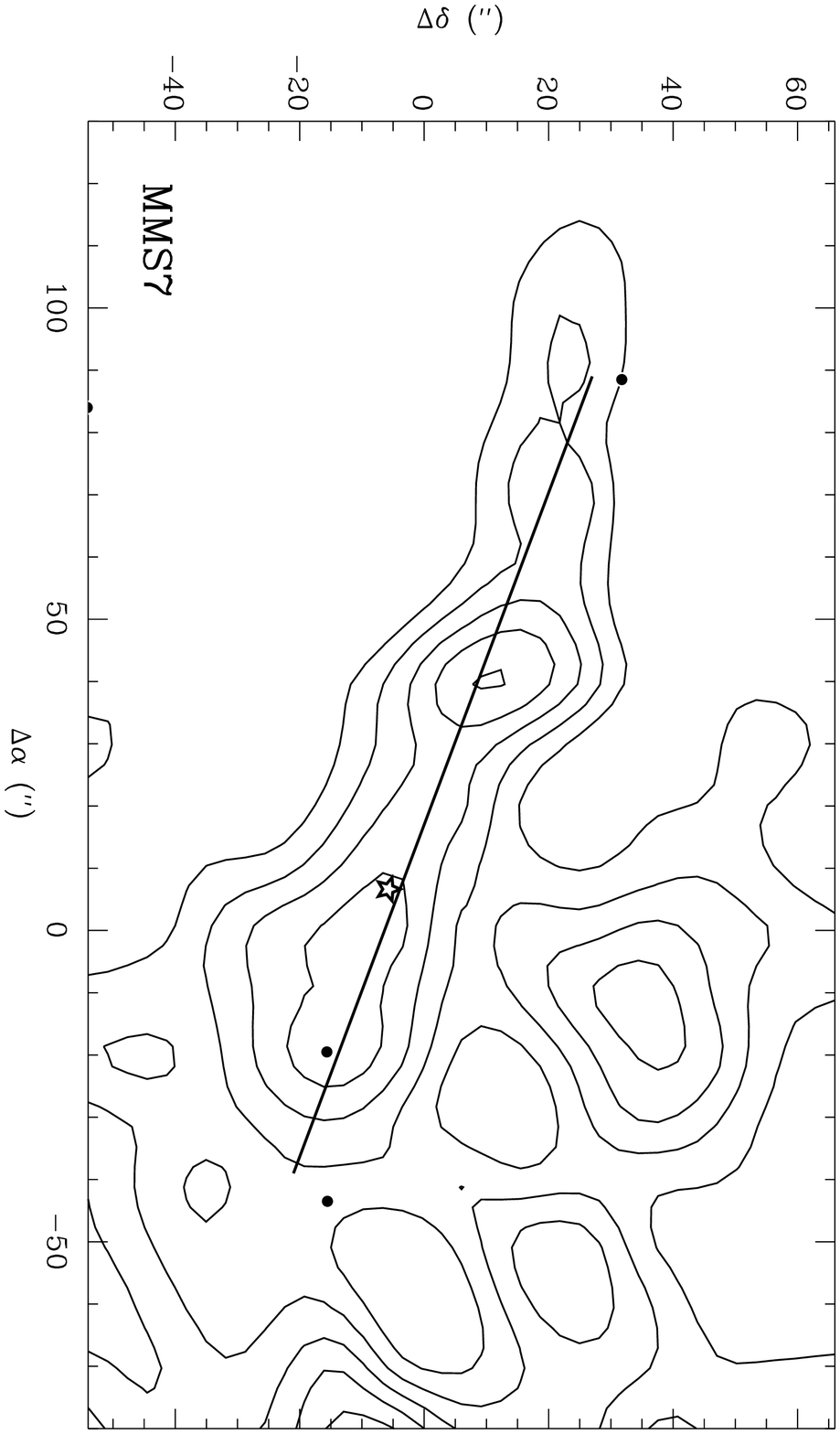}{50pt}{90}{60}{60}{240}{-300}
\end{figure}
\vskip 3.2in
\startfigcap
\noindent{\bf Figure 4:}
Contours of BIMA+FCRAO integrated emission from $v=11.5$ to 13.5~\kms\
in the region around MMS7. Because the velocity of this flow is similar
to that of the bulk of the cloud emission the total emission is very strong
and contouring begins at 33~K~\kms. The contour step is 3~K~\kms.
The solid dark line schematically marks the position of this flow
but no velocity gradient across it could be reliably measured.
\endfigcap

\clearpage
\begin{figure}[ht]
\plotfiddle{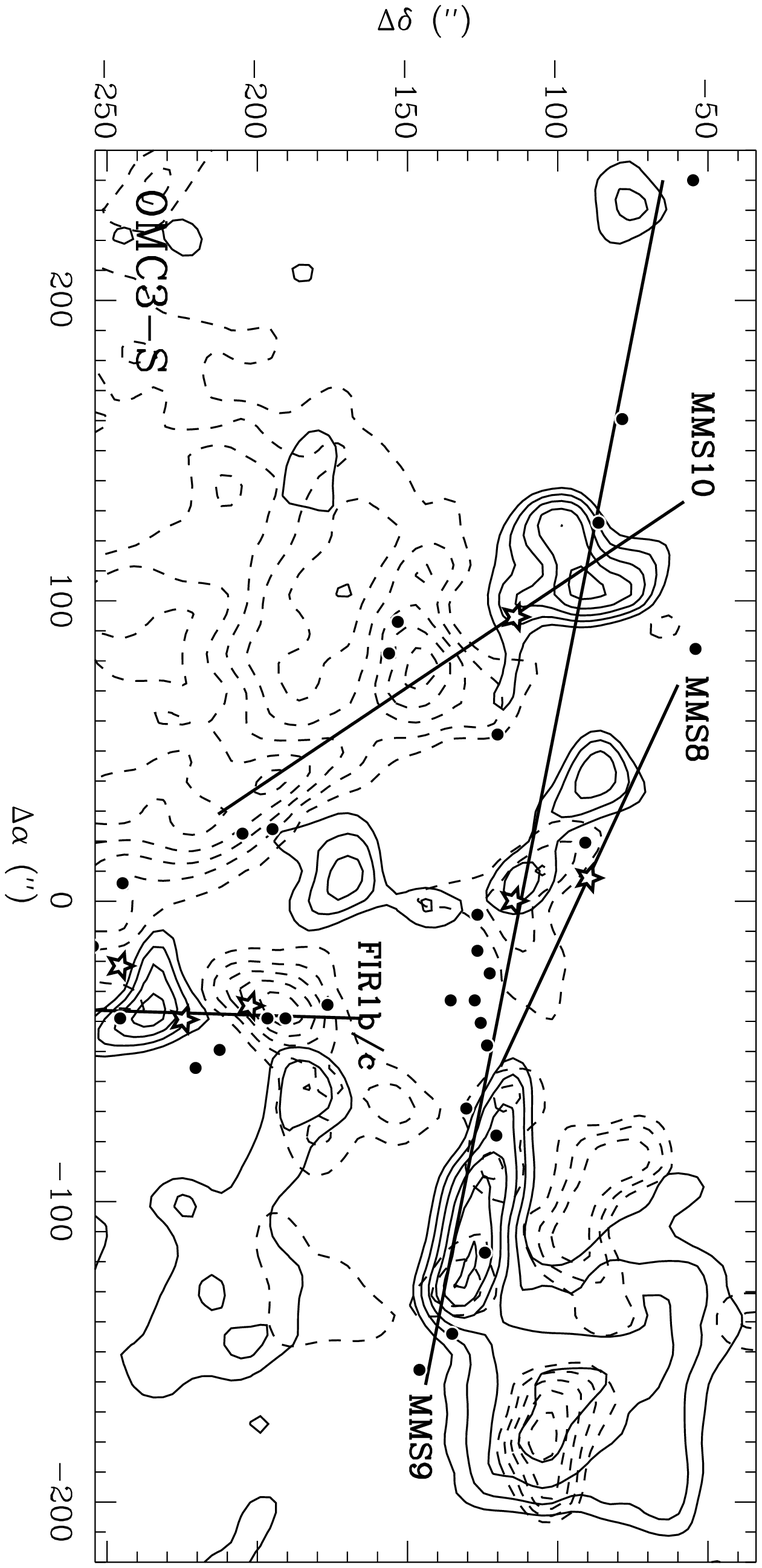}{50pt}{90}{60}{60}{240}{-300}
\end{figure}
\vskip 3.0in
\startfigcap
\noindent{\bf Figure 5:}
As Figure 3 for the southern part of OMC-3 containing sources MMS8-10.
Dashed contours (colored blue in electronic edition) are for
$v=4.0$ to 8.0~\kms\ and solid contours (colored red in electronic edition)
are for $v=13.0$ to 16.0~\kms. Contours begin at 12~K~\kms\ and
increment at 4~K~\kms.
\endfigcap

\clearpage
\begin{figure}[ht]
\plotfiddle{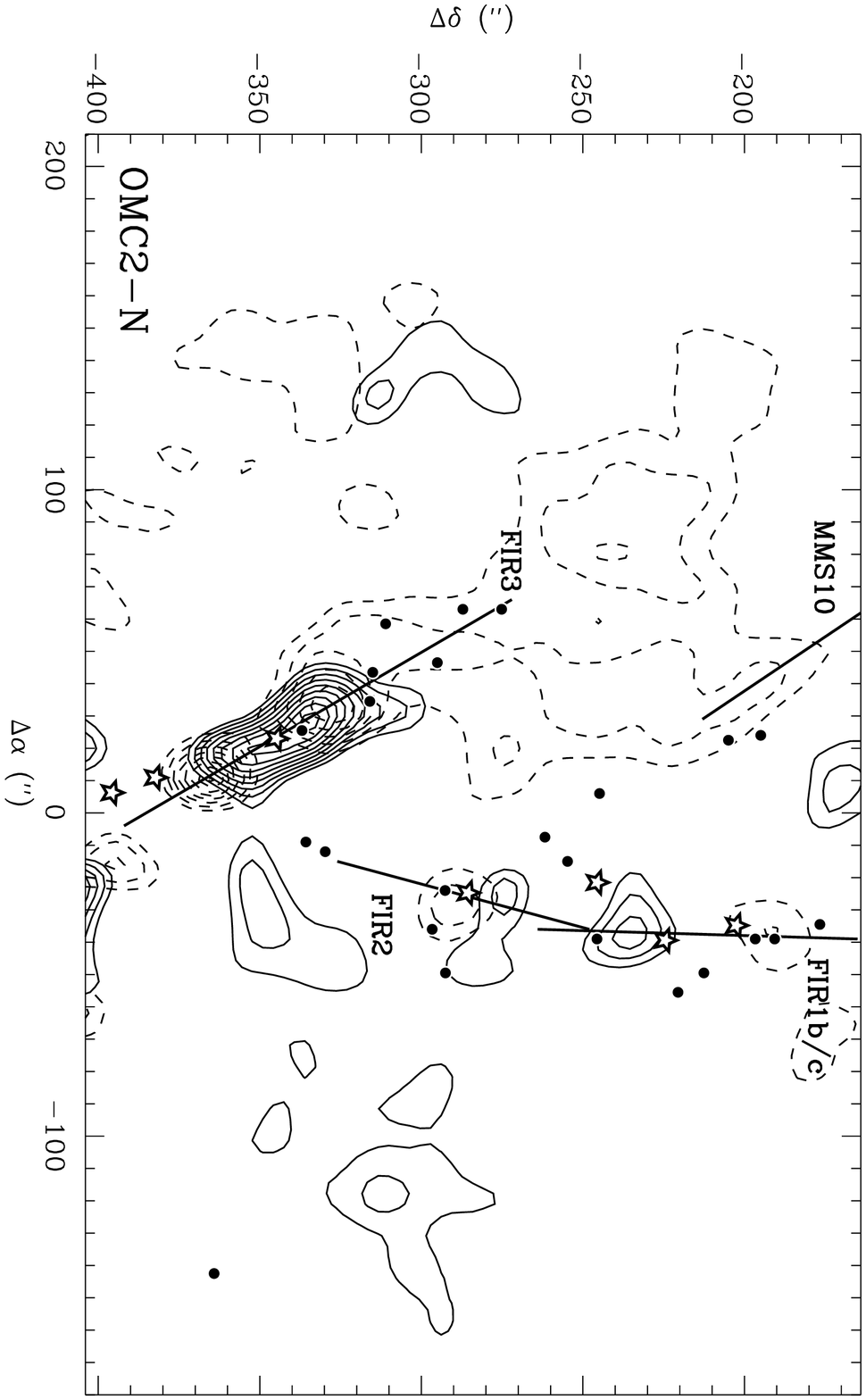}{50pt}{90}{60}{60}{240}{-300}
\end{figure}
\vskip 3.3in
\startfigcap
\noindent{\bf Figure 6:}
As Figure 3 for the northern part of OMC-2 containing sources FIR1-6.
Dashed contours (colored blue in electronic edition) are for
$v=4.0$ to 7.5~\kms\ and solid contours (colored red in electronic edition)
are for $v=13.5$ to 17.0~\kms. Contours begin at 12~K~\kms\ and
increment at 4~K~\kms.
\endfigcap

\clearpage
\begin{figure}[ht]
\plotfiddle{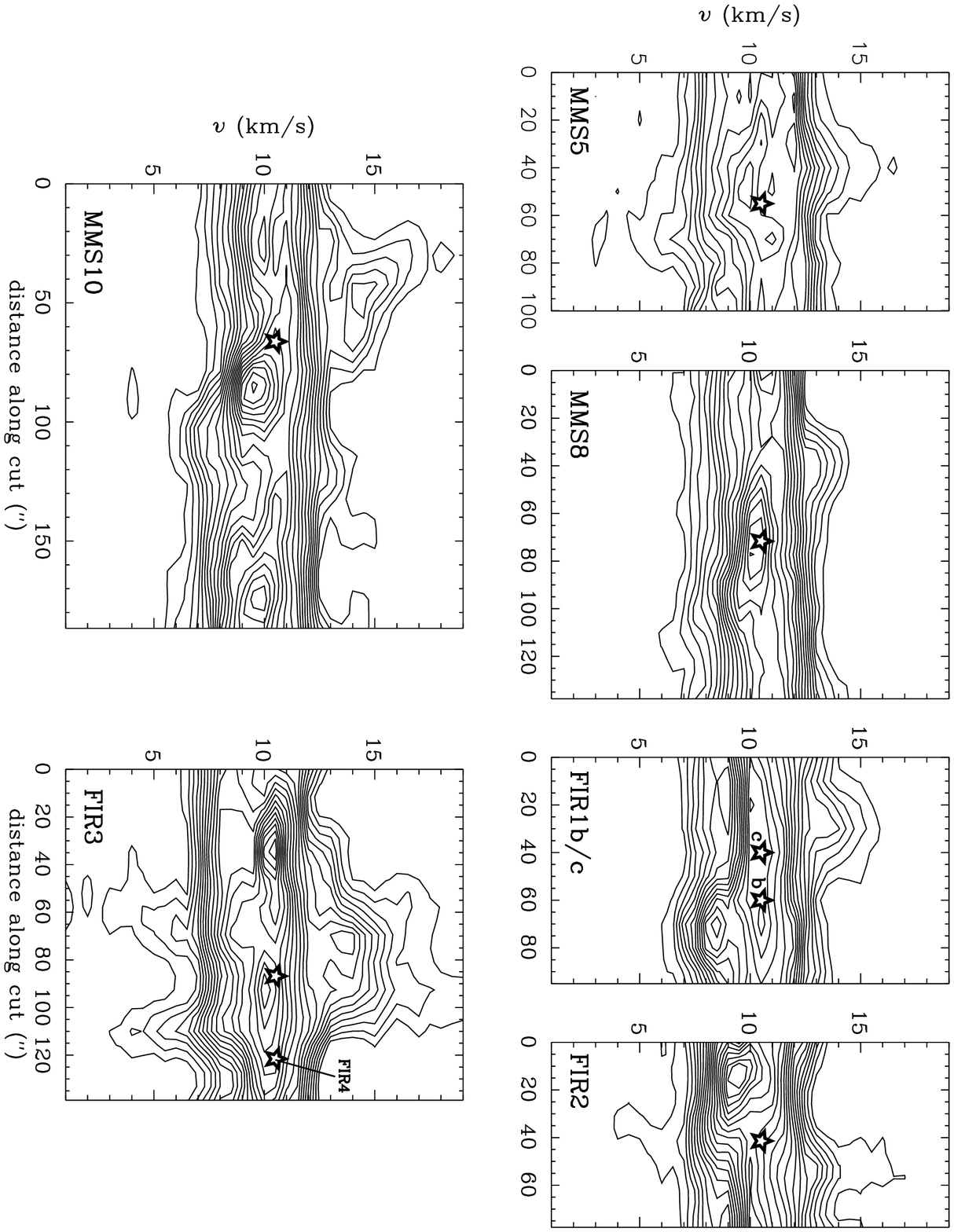}{50pt}{90}{65}{65}{265}{-320}
\end{figure}
\vskip 3.8in
\startfigcap
\noindent{\bf Figure 7:}
Position-velocity maps along the cuts defined in Figures 3, 5, 6.
The position increases from the leftmost (east) portion of the cut
in each case. The long outflows from MMS2/3 and MMS9 that lie
predominantly along the plane of the sky have low velocity gradients
and are are not shown. The contour starting
level and increment are 2~K for all maps. The location of the
Chini et al. protostars along the cut (at an assumed $v=10.5$~\kms)
are shown by the star symbols and labeled in the two cases where more than
one source lies along the cut.
\endfigcap

\clearpage
\begin{table}
\begin{center}
TABLE 1\\
Outflow properties\\
\vskip 2mm
\begin{tabular}{lccccc}
\hline\\[-2mm]
 & $M$ & $P$ & $E$ & $L$ & $dV/dR$ \\[-1mm]
 & ($M_\odot$) & ($M_\odot$~\kms) & ($10^{43}$ erg) & (pc) & (\kms pc\e) \\[2mm]
\hline \hline \\[-3mm]
MMS2/3  &&&&&\\[-1mm]
blue lobe   & 0.13  &  0.38  &  1.26 &&\\[-1mm]
red lobe    & 0.08  &  0.24  &  0.78 &&\\[-1mm]
total       & 0.21  &  0.62  &  2.04  &    0.54  &  21 \\[1mm]
MMS5    &&&&&\\[-1mm]
blue lobe   & 0.03  &  0.10  &  0.40 &&\\[-1mm]
red lobe    & 0.02  &  0.09  &  0.34 &&\\[-1mm]
total       & 0.05  &  0.19  &  0.74  &    0.13  & 100 \\[1mm]
MMS7    &&&&&\\[-1mm]
plane of sky & 0.04  &        &        &    0.38  &     \\[1mm]
MMS8    &&&&&\\[-1mm]
blue lobe   & 0.02  &  0.07  &  0.23 &&\\[-1mm]
red lobe    & 0.04  &  0.13  &  0.44 &&\\[-1mm]
total       & 0.06  &  0.20  &  0.67  &    0.20  &  55 \\[1mm]
MMS9    &&&&&\\[-1mm]
plane of sky & 0.22  &        &        &    0.87  &     \\[1mm]
MMS10   &&&&&\\[-1mm]
blue lobe   & 0.16  &  0.55  &  2.08 &&\\[-1mm]
red lobe    & 0.14  &  0.49  &  2.04 &&\\[-1mm]
total       & 0.30  &  1.04  &  4.12  &    0.39  &  34 \\[1mm]
FIR1b/c &&&&&\\[-1mm]
blue lobe   & 0.11  &  0.28  &  0.79 &&\\[-1mm]
red lobe    & 0.05  &  0.16  &  0.52 &&\\[-1mm]
total       & 0.16  &  0.44  &  1.31  &    0.16  &  66 \\[1mm]
FIR2    &&&&&\\[-1mm]
blue lobe   & 0.02  &  0.08  &  0.29 &&\\[-1mm]
red lobe    & 0.03  &  0.10  &  0.39 &&\\[-1mm]
total       & 0.05  &  0.18  &  0.68  &    0.08  & 190 \\[1mm]
FIR3$^\ast$  &&&&&\\[-1mm]
blue lobe   & 0.12  &  0.42  &  1.67 &&\\[-1mm]
red lobe    & 0.21  &  0.74  &  3.32 &&\\[-1mm]
total       & 0.33  &  1.16  &  4.99  &    0.27  & 66 \\[2mm]
\hline\\[-4mm]
\multicolumn{6}{l}{$^\ast$Appears to be the sum of two overlapping flows.}
\end{tabular}
\end{center}
\end{table}

\end{document}